\documentclass[11pt,twoside]{article}
\usepackage{RAFproc}

\resetcounters
\begin{document}

\title{How shall we determine detection sensitivity in radio pulsar search?}

\author{Meng Yu$^{1,2}$ \affil{$^1$Joint Laboratory for Radio Astronomy
    Technology, National Astronomical Observatories, Chinese Academy
    of Sciences, Beijing, China} \affil{$^2$CAS Key Laboratory of
    FAST, National Astronomical Observatories, Chinese Academy of
    Sciences, Beijing, China; \email{meng.yu@nao.cas.cn}}}

\begin{abstract}
Determination of detection sensitivity in a number of previous pulsar
search programmes was done via the straightfoward use of the
radiometer equation. In the same surveys, the Fourier domain method
was used to search for pulsars. As detection sensitivity is partially
a function of the searching method, the straightfoward use of the
radiometer equation for detection sensitivity determination is not
consistent with the Fourier searching method. In this proceeding, I
clarify the problem and note the way for sensitivity determination
which is consistent with the Fourier searching method. More details
can be found in \citet{yu18}.
\end{abstract}

\section{The problem}
To determine the detection sensitivity is an essential requirement of
a pulsar search programme. In the Princeton-NRAO (phase I) pulsar
survey, \citet{dtws85} used the radiometer equation to determine the
survey sensitivity. In their implementation, they first presented the
form of the equation derived by introducing the top-hat pulse signal
(see Eq. 1 in \citealt{dtws85} or Eq. A1.22 in \citealt{lk05}), then
they set the integration time in the equation to be the entire
integration time per telescope pointing and set the confidence limit
as 7.5\,$\sigma$ via the pulse signal-to-noise. The pulse
signal-to-noise is defined as the proportion of height of the top-hat
profile to the standard deviation of the profile baseline.

The \citet{dtws85} method has subsequently been used for the
high-frequency southern Galactic plane pulsar survey \citep{jlm+92},
the Parkes multi-beam pulsar survey \citep{mlc+01} and the PALFA
survey \citep{cfl+06} etc. The southern Galactic plane survey had
integration time 78.6\,s or 157.3\,s, sampling time 0.3\,ms or 1.2\,ms
and the confidence limit 8.0\,$\sigma$, the multi-beam survey had
integration time 35\,min, sampling time 0.25\,ms and the confidence
limit 8.0\,$\sigma$ and the PALFA survey had integration time 134\,s
or 67\,s, sampling time 1.024\,ms and the confidence limit
10.0\,$\sigma$. However, in the data processing for these surveys, the
Fourier domain methods were used. The idea of the Fourier domain
searching methods is to make use of the high sensitivity of the
Fourier transform to periodicity. In the methods, a power or an
amplitude spectrum is firstly derived by Fourier transforming an
observed and dedispersed time series. Then a detection threshold is
drawn out of the spectra probability distribution and the independent
spectra number searched under a given confidence level. Finally pulsar
candidates are those spectra with heights higher than the
threshold. Therefore, on the determination of survey sensitivity, the
\citet{dtws85} method is not consistent with the Fourier domain
searching methods and should not be used for the search programmes
that use the Fourier domain methods to search for pulsars.

\section{The solution}

\begin{figure}
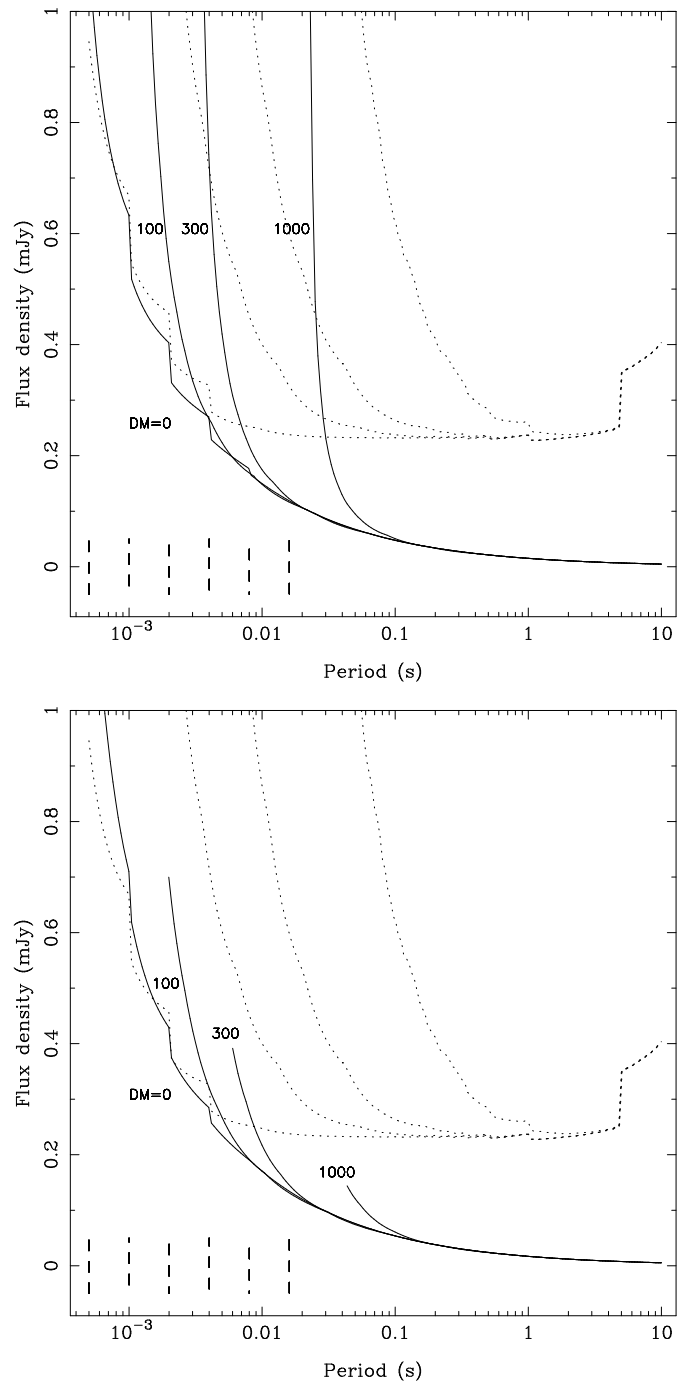

\begin{center}
\begin{tabular}{c}
\includegraphics[width=9cm,angle=-90]{flux_sq_2.eps}\\
\includegraphics[width=9cm,angle=-90]{flux_vm_2.eps}
\end{tabular}
\caption{Minimum detectable mean flux densities derived in my trial
  (solid lines) or via the original routine (dashed lines) for the
  Parkes multi-beam pulsar survey. The upper panel is for the top-hat
  profile model and the lower panel is for the modified von Mises
  profile model.}\label{fig:flux}
\end{center}
\end{figure}

For the Fourier domain searching methods, \citet{vvw+94} have given
the definition of sensitivity and the routine for deriving it. The
point lies in establishing the relation between a pulse profile and
the spectra threshold. They presented implementation of their routine
with a sinusoidal profile. That was for their X-ray pulsar search
programme. For radio pulsar search, the \citet{vvw+94} routine can be
implemented with the following procedures:
\begin{description}

\item[i] Under a given confidence level, derive the Fourier domain
  detection threshold with the spectra probability distribution and
  the number of independent spectra searched. As harmonic summing
  techniques are normally used in radio pulsar search, detection
  thresholds of the folded spectra should also be derived;

\item[ii] Under a given confidence level, derive powers of signal
  corresponding to the detection thresholds with the noise-signal
  probability distribution \citep{gro75d};

\item[iii] Setup the relations between the signal powers and pulse
  profiles. Then deduce the amplitudes of the pulses via the
  relations. The derived amplitudes are the minimum detectable mean
  amplitudes, or sensitivities, in the arbitrary unit;

\item[iv] Convert the derived amplitudes into flux densities via the
  radiometer equation.
  
\end{description}
I have attempted the procedures; details are given in \citet{yu18}. In
my trail, the top-hat and the modified von Mises \citep{rem02} profile
models were used. When converting the amplitudes into flux densities,
the published system parameters of the Parkes multi-beam pulsar survey
\citep{mlc+01} were used. In Fig. \ref{fig:flux}, the derived
sensitivities as a function of pulsar period with dispersion measure
${\rm DM} = 0, 100, 300$ and 1000\,cm$^{-3}$\,pc are shown. The
sensitivities given by the routine which was developed by
\citet{cra00} as the orignal sensitivity predictions are also
shown. We see there are wide discrepancies between sets of the
curves. This is primarily because the sensitivities given in my trial
are drawn out of the 3\,$\sigma$ confidence limit, while those of the
original predictions are drawn out of the 8\,$\sigma$ confidence
limit.

\acknowledgements I acknowledge the anonymous referee for
\citet{yu18}, who has given the comments that have greatly helped in
improving the manuscript. I acknowledge Prof. F. Crawford for the
helpful discussion with him and his provision of the original routine
for the sensitivity prediction of the Parkes multi-beam pulsar
survey. This work is supported by 1) the National Key Project of
China, Frontier research on radio astronomy technology; 2) the Open
Project Program of the Key Laboratory of FAST, National Astronomical
Observatories, Chinese Academy of Sciences; 3) the National Natural
Science Foundation of China (No. 11261140641, No. 11403060).

\bibliographystyle{RAFproc}
\bibliography{psrrefs,modrefs,crossrefs,journals}

\end{document}